\documentclass{aa}  

\usepackage{graphicx}  
\usepackage[varg]{txfonts}
\usepackage{hyperref}
\hypersetup{
    colorlinks=true,
    linkcolor=blue,
    filecolor=magenta,      
    urlcolor=cyan,
   citecolor=blue,
}
\defcitealias{2020kri}{Paper I}
\begin{document}

   \title{Temperature diagnostics of chromospheric fibrils}

   \author{M. Kriginsky\inst{1,2} \and R. Oliver\inst{1,2} \and D. Kuridze\inst{3,4}}

   \institute{Departament de F\'\i sica, Universitat de les Illes Balears, E-07122 Palma de Mallorca, Spain
         \and
          Institute of Applied Computing \& Community Code (IAC3), UIB, Spain
            \and
            Department of Physics, Aberystwyth University, Ceredigion, SY23 3BZ, UK 
          \and
            Abastumani Astrophysical Observatory, Mount Kanobili, 0301, Abastumani, Georgia
                      }

   \date{Received ; accepted }

 
  \abstract
   {Chromospheric fibrils are thin and elongated structures that connect nearby photospheric magnetic field concentrations of opposite polarities.}
   { We assess the possibilities and drawbacks related to the use of current instrumentation and inversion techniques to infer the thermodynamic structure of chromospheric fibrils.}
   {We employed spectroscopic observations obtained in the \ion{Ca}{II} 854.2~nm line with the CRISP instrument at the Swedish 1-m Solar Telescope and in coordination with observations in the ultraviolet \ion{Mg}{II} h \& k lines taken with the IRIS satellite. We studied the temperature sensitivity of these chromospheric lines to properly invert their spectral profiles with the Stockholm inversion Code and determine the temperature, line-of-sight velocity, and microturbulent velocity of manually traced chromospheric fibrils present in the field of view.}
   {Fibril-like structures show a very particular dependence of their temperature as a function of the position along their length. Their temperatures at the detected footpoints are, on average, 300 K higher than the temperature at the midpoint. The temperature variation appears to be almost symmetrical in shape, with partially traced fibrils showing a similar trend for the temperature variation. Additionally, the response of the \ion{Ca}{II} 854.2~nm line core to variations of the temperature for the inverted models of the atmosphere in fibril areas seems to be insufficient to properly resolve the aforementioned temperature structure. Only  the addition of more temperature sensitive lines such as the \ion{Mg}{II} h \& k lines would make  it possible to properly infer the thermodynamic properties of chromospheric fibrils. Comparisons between the results obtained here and in previous studies focused on bright \ion{Ca}{II} K fibrils yield great similarities between these structures in terms of their temperature. } 
   {}


   \titlerunning{}
   \authorrunning{M. Kriginsky et al.}
   \maketitle
   
%

\section{Introduction}

The dynamic and ever-changing nature of the solar chromosphere spawns a myriad of structures. One of the most notable types of structures are the thin, dark, and elongated features that appear to connect neighbouring magnetic field concentrations of opposite polarities.  Referred to as chromospheric fibrils, such structures are thought to be closely aligned with low-lying, nearly horizontal magnetic fields. This assumption has been addressed in several works, both based on spectropolarimetric observations \citep[][]{2011A&A...527L...8D,2013ApJ...768..111S,2015SoPh..290.1607S,2017A&A...599A.133A,2017ApJS..229...11J} and via analyses of numerical simulations \citep[][]{2015ApJ...802..136L,2016ApJ...831L...1M}. The general consensus is that fibrils do indeed appear to be aligned with the magnetic field orientation, as traditionally believed; however, there are cases where this alignment is not present.

 The fibril-like phenomena that populate the chromosphere observed on the disc and off the solar limb are also largely seen to show oscillatory motions \citep[see e.g.][]{2012ApJ...750...51K}, a clear signature of propagating waves along their length \citep[see review by][]{2009SSRv..149..355Z}, which can contribute to the heating of the plasma in these structures. In fact, numerical models of wave propagation and dissipation are often used to infer properties of the plasma \citep[see e.g. ][]{2017A&A...607A..46M}.

 \begin{figure*}
   \centering
   \includegraphics[width=8cm,height=8cm]{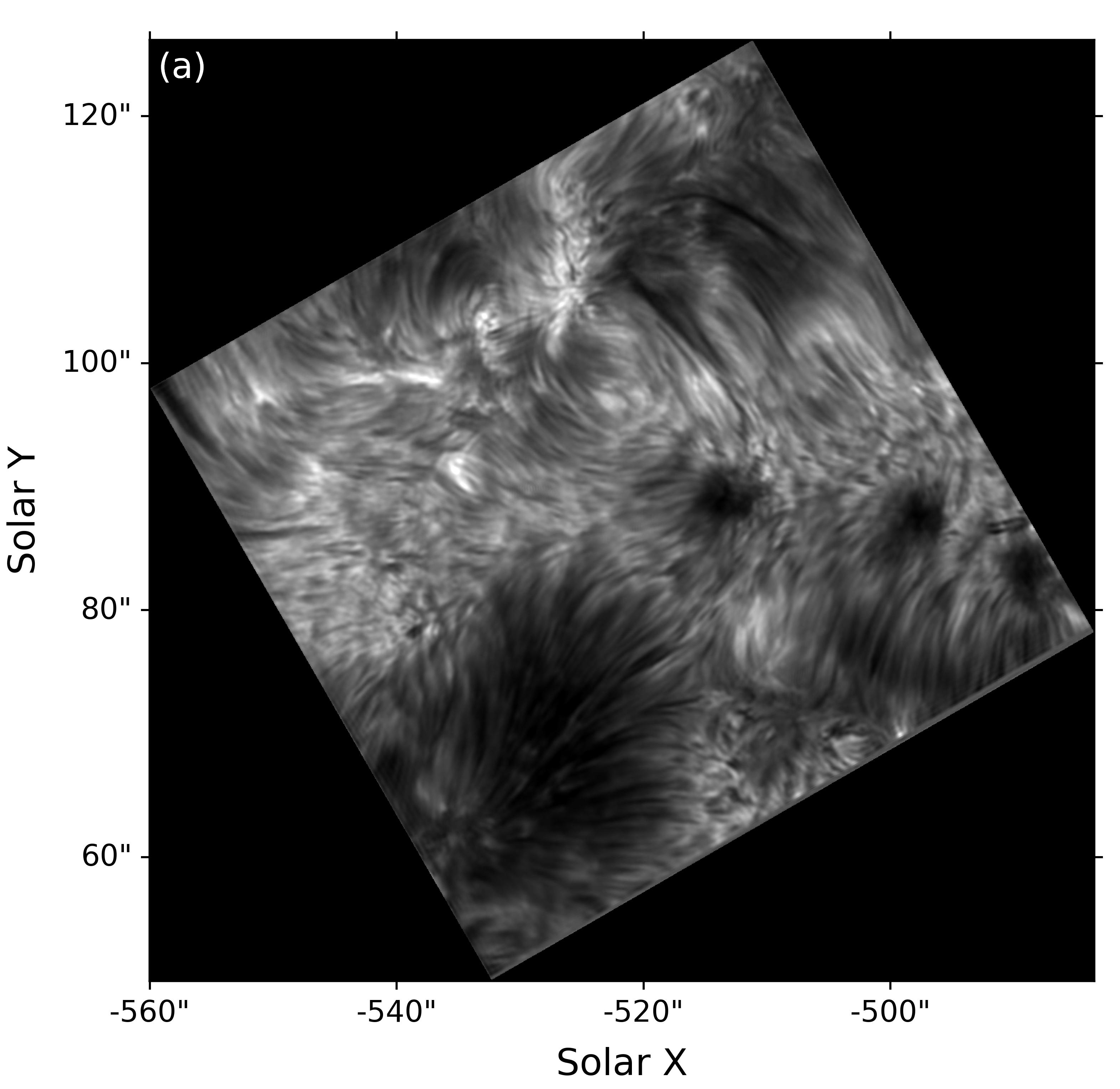}
    \includegraphics[width=10.16cm,height=8cm]{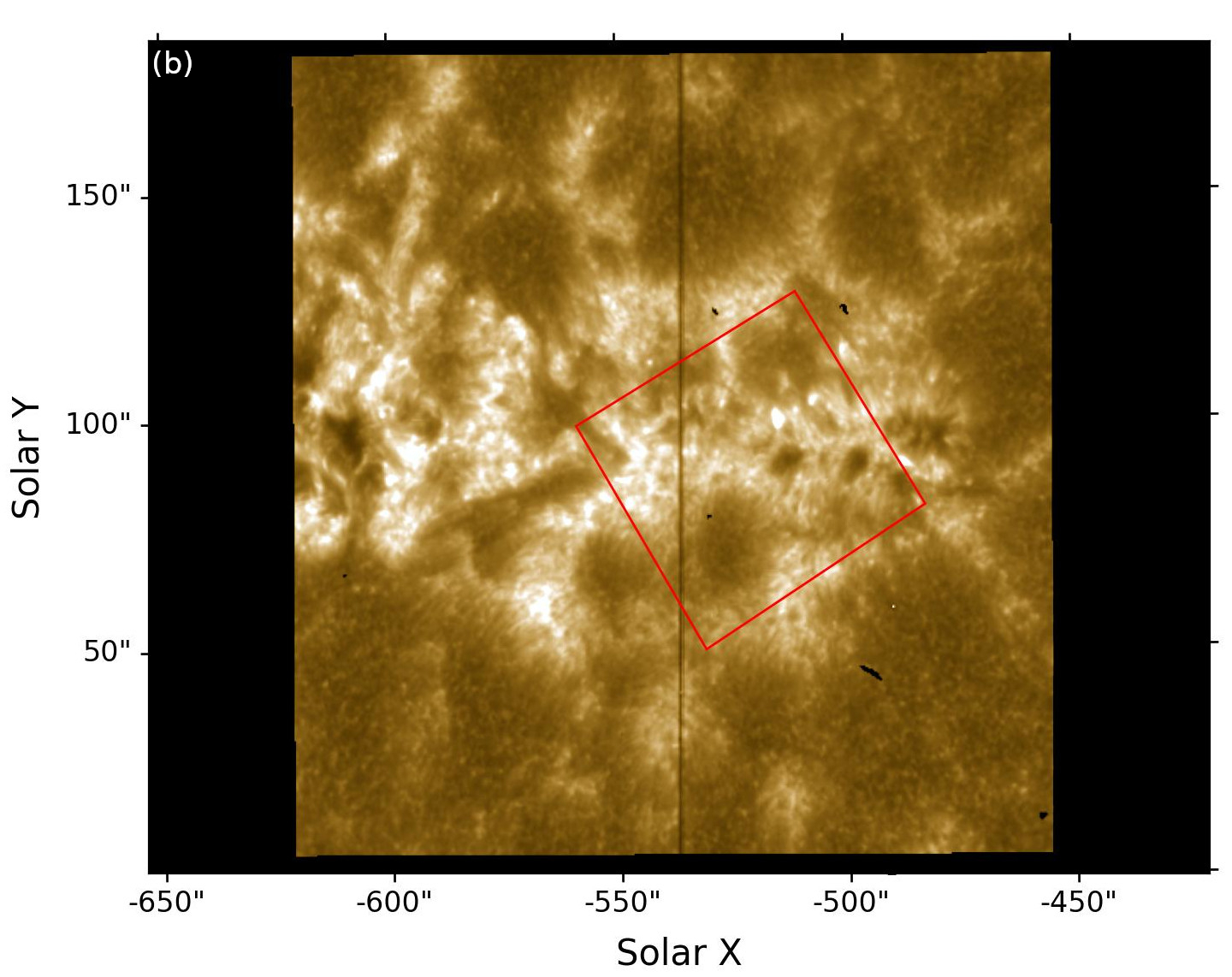}
     
   \caption{Context of the observations. (a)  CRISP field of view at the core of the \ion{Ca}{II} 854.2~nm line at 07:55:21 UT and (b) Its superposition outlined with a red rectangle over the IRIS 279.6~nm slit-jaw passband at 07:55:12 UT.}
              \label{Figure:1}%
\end{figure*}

Determining the properties of chromospheric fibrils is of paramount importance in correctly modelling and understanding the mass and energy flows that take place in the solar atmosphere. Part of this process must come from inferring physical quantities such as temperature, magnetic fields, and velocity fields from direct observations. This is a non-trivial problem in the case of the chromosphere, since the plasma density falls precipitously with height, causing the collisional rates to drop substantially. Line formation in these circumstances is much more susceptible to the properties of the radiation field than it is in the photosphere. The problem then acquires a non-local nature, where the properties of the plasma in a position in space become dependent on the conditions elsewhere in the atmosphere. Recent advances in the field of non-local thermodynamical equilibrium (non-LTE) inversions of the radiative transfer equation \citep[for a review, see][]{2016LRSP...13....4D} have made it possible to obtain the stratification of different physical quantitities from spectroscopic and spectropolarimetric observations of the chromosphere.

 The Stockholm inversion code \citep[STiC;][]{2016ApJ...830L..30D,2019A&A...623A..74D} is one such inversion code. Considering line formation in the Zeeman regime, STiC is able to perform inversions under non-LTE conditions of multiple spectral lines from different atoms at the same time. This opens the door to the simultaneous inversion of data obtained from different instruments sampling multiple regions of the solar atmosphere, offering the possibility to better constrain the inverted physical parameters \citep[][]{2018A&A...620A.124D}.
 
As summarised by \citet{2012SSRv..169..181T}, there have been several attempts at determining the temperature and other spectroscopic properties of on-disc fibrillar structures. Using contrast non-LTE H$\alpha$ line observations, \citet[][]{1967AuJPh..20...81G,1967sp...conf..353G} concluded that dark, fibrillar chromospheric structures exhibit temperatures no higher than $10^4$~K. Another widely used method is the so-called 'cloud' model \citep[][]{1964PhDT........83B,1990A&A...230..200A,2012SSRv..169..181T}, which allows for the determination of four adjustable parameters: the Doppler width, $\Delta\lambda_{\mathrm{D}}$, the optical thickness, $\tau_0$, the cloud's line-of-sight velocity, $v,$ and the source function, $S$. \citet[][]{1997A&A...324.1183T} devised a method that uses the four aforementioned parameters in order to determine other physical quantities \citep[see][ for a detailed description]{1997A&A...324.1183T,2012SSRv..169..181T}. From the determination of the Doppler width and assuming non-thermal velocities of 10 and 15~km~$\mathrm{s}^{-1}$, \citet[][]{1997A&A...324.1183T} inferred temperatures of $1.4 \times 10^4$~K and $1.0 \times 10^4$~K, respectively. More recently, \citet[][]{2020A&A...637A...1K} studied the so-called \ion{Ca}{II} K bright fibrils using spectroscopic observations in the \ion{Ca}{II} K line, along with spectropolarimetric observations in the \ion{Ca}{II} 854.2~nm and \ion{Fe}{I} 630.2~nm lines. These authors performed non-LTE inversions of the bright fibrillar material, with temperatures on the order of 5000 to 6000~K  recovered from the inversions, representing a lower value than that obtained with cloud modelling of dark fibrillar material.

In this paper, spectroscopic observations obtained with the Swedish 1-meter Solar Telescope \citep[SST;][]{sst} of a plage region containing low-lying dark chromospheric fibrils rooted in magnetic field concentrations were used  to study their thermodynamic properties. Accompanying spectra obtained with the Interface Region Imaging Spectrograph \citep[IRIS;][]{2014SoPh..289.2733D} were used together with the SST data to infer the thermal stratification of the solar atmosphere in these structures using multi-line inversions. Section~\ref{sec:obs} offers a description of the observations and data alignment. In Sect.~\ref{sec:Data}, we describe the inversion strategy. Section~\ref{results} presents the results, with Sect.~\ref{conclusions} presenting our final conclusions.

\begin{figure*}
 \centering
 \includegraphics[width=18cm]{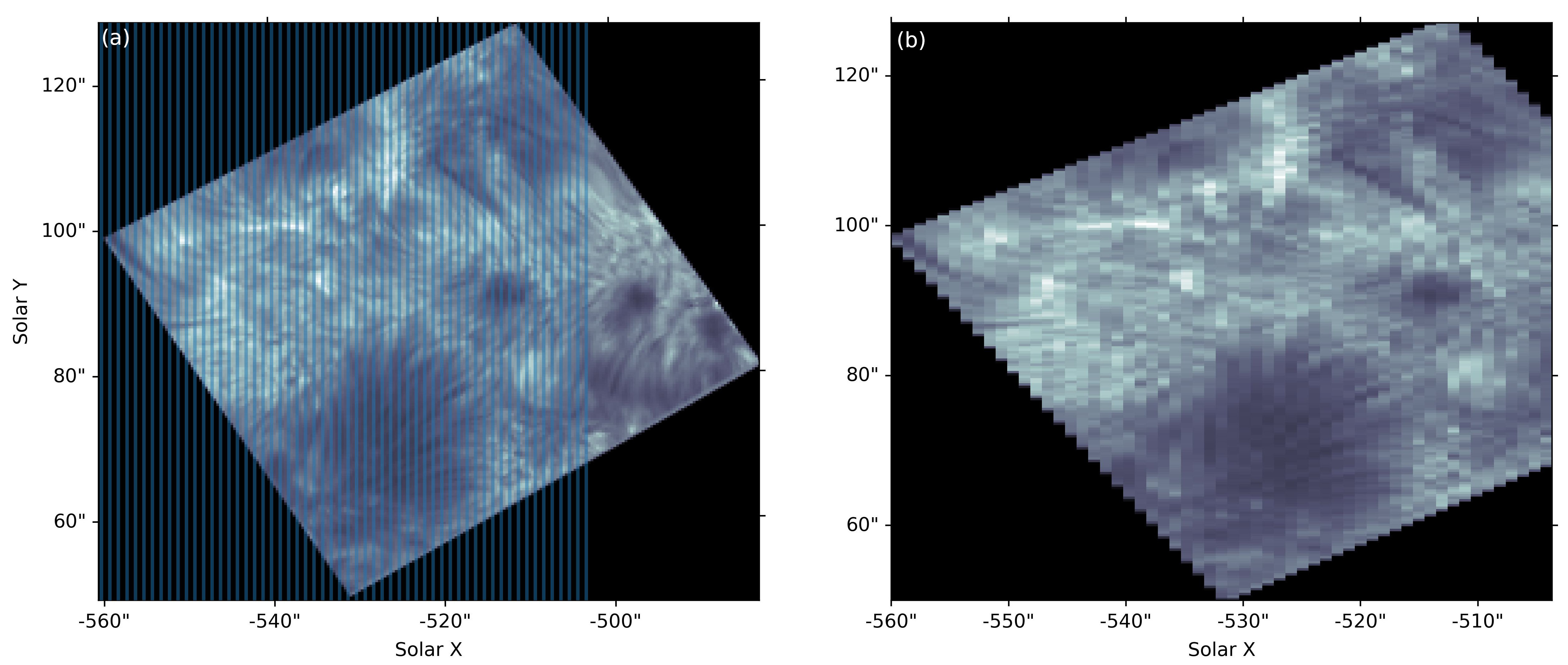}
 \caption{Spatio-temporal alignment example. (a) Spatially averaged and rotated \ion{Ca}{II} line core image with the vertical blue lines representing the position of the IRIS slit over time and (b) \ion{Ca}{II} line core re-sampled to the IRIS slit positions. }
 \label{Figure:2}%
\end{figure*}

 \section{Observations} \label{sec:obs}

The active region NOAA 12661 was centred at solar coordinates ($X$,$Y$)= (-570\arcsec, 100\arcsec) in 04 June 2017. The western side of the active region was the subject of a coordinated observation campaign between IRIS and the SST (Fig. \ref{Figure:1}).

\subsection{SST data}
Spectroscopic data centred at heliocentric angle $\mu = 0.87$ were obtained with the  CRisp Imaging SpectroPolarimeter \citep[CRISP;][]{Scharmer_2008} in the \ion{Ca}{II} 854.2~nm (hereafter also referred to as the \ion{Ca}{II} line)  and H$\alpha$ lines. The calcium line was sampled over 27 spectral positions spanning the wavelengths at $\pm{2}$~\AA\ from the line core in a non-uniform spectral grid that was denser at the line core ($\Delta \lambda = 0.05$~\AA) and sparser at the line wings ($\Delta \lambda = 0.25$~\AA). The H$\alpha$ line was spanned with an identical wavelength sampling, also with 27 spectral positions. The total duration of the observations was 36~min, with a cadence of approximately 15~s, totalling 136 scans. 

Reconstruction of the data was done with the use of the Multi-Object Multi-Frame Blind Deconvolution \citep[MOMFBD;][]{VanNoort2005} method. The CRISP data reduction pipeline \citep[CRISPRED;][]{2015delacruz} was also used for additional data reduction, including the cross-correlation method of
\citet{Henriques2012A}. An absolute wavelength calibration of the \ion{Ca}{II} line was done using the atlas profile of \citet[][]{1984SoPh...90..205N}, accounting for limb darkening effects.

\begin{figure*}
   \centering

   \includegraphics[width=18.16cm]{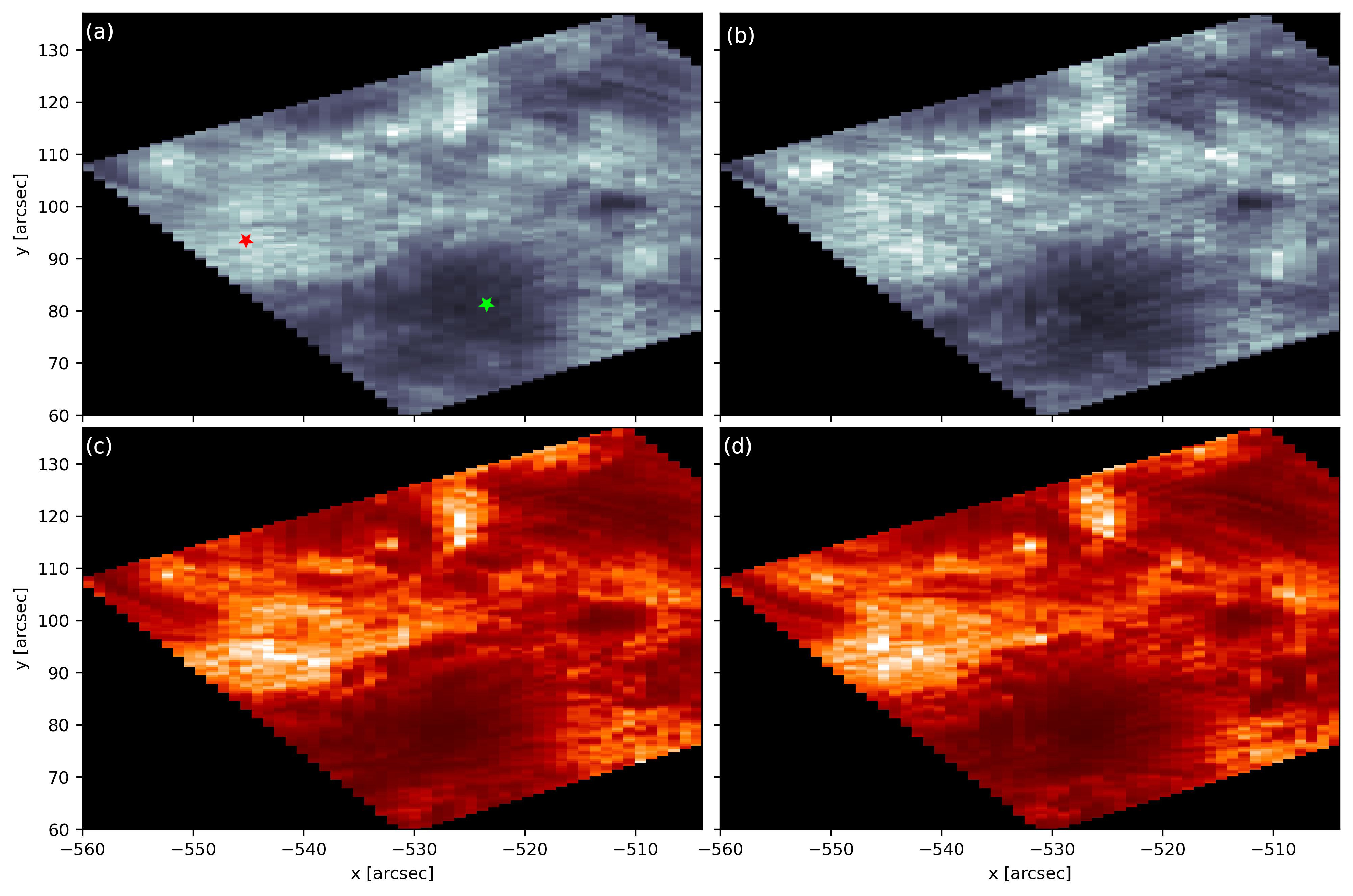}
   \caption{Aligned \ion{Ca}{II} and \ion{Mg}{II} observations. \ion{Ca}{II} line core images for (a) sets A and (b) B, and \ion{Mg}{II} k line core images also for (c) sets A and (d) B. The brightness of each image has been scaled independently. The red star in panel (a) marks the location of the representative pixel used in Fig.~\ref{Figure:6}, while the green star marks the location of the pixel used in Fig.~\ref{Figure:4}.}
              \label{Figure:3}%
\end{figure*}
\subsection{IRIS data}
IRIS support for this observation consisted of very large sparse 64-step raster scans (OBSID 3620110460). The raster scan covered a field of view of 64\arcsec x175\arcsec, with non-contiguous 1\arcsec steps with a slit width of 0\arcsec .33. The exposure time at each slit position was roughly 16~s, resulting in an overall cadence of 17 min per raster scan. A total of 11 complete raster scans were obtained, but only the first two were used in this study as they were the ones which corresponded to the timeline of the SST observations. Hereafter, they will be referred to as sets A and B. Context images from the 276.2~nm, 283.2~nm and \ion{Si}{IV} slit-jaw passbands were also obtained.

The IRIS data calibration was done following the reduction methods described in detail in \citet[][]{2018SoPh..293..149W}. Radiometric calibration of the raster data was carried out using version four of the calibration files employed by the
\textit{iris\_get\_response} routine in SolarSoft \citep[SSW;][]{1998SoPh..182..497F}.

\subsection{Data alignment}

Alignment between the CRISP and IRIS data was done by cross-correlating the slit-jaw 283.2~nm images with the far wing images of the \ion{Ca}{II} data at the time of best seeing conditions in the SST. After alignment, the CRISP observations were re-sampled to match the spatial sampling of the IRIS raster observations. This spatial averaging was done in order to resolve the same features with the same spatial resolution in both instruments. Therefore, there was a one-to-one correspondence between IRIS and SST pixels. The observations at the original resolution of the SST were obtained at sub-optimal seeing conditions, therefore a spatial averaging also helped to improve the signal-to-noise ratio of the data.

Aligning raster data from IRIS and data from CRISP meant dealing with the alignment of a spectrograph and a filtergraph. In order to maximise the spatial alignment while minimising the time difference between the observations from both instruments for each pixel, the CRISP data were taken at the positions of the IRIS raster scan with the minimum time delay between both exposures, thus effectively creating a spectrograph slit observation for the CRISP data (see Fig.~\ref{Figure:2}). Figure \ref{Figure:3} shows the results of the alignment for sets A and B. With this alignment strategy, the maximum time difference between a co-spatial \ion{Ca}{II} pixel and a \ion{Mg}{II} pixel was 7.11~s. The 0\arcsec.67 gaps between consecutive horizontal pixels resulting from the IRIS data acquisition set-up were not interpolated in order to fill the horizontal field of view, since this would have led to a data set containing twice as many interpolated pixels than observed pixels.

 \section{Inversions} \label{sec:Data}
 
 \subsection{STiC}
 The inversion of the CRISP and IRIS data was performed using the MPI-parallel non-LTE code STiC -- a regularised Levenberg-Marquardt code that uses an optimised version of the RH code \citep[][]{2001ApJ...557..389U} to solve the atom population densities assuming statistical equililibrium and plane-parallel geometry for multi-atom non-LTE inversions of multiple spectral lines. It allows for the inclusion of partial redistribution effects of scattered photons \citep[][]{2012A&A...543A.109L}. The radiative transport equation is solved using cubic Bezier solvers \citep[][]{2013ApJ...764...33D}. The inversion engine of STiC includes an equation of state extracted from the Spectroscopy Made Easy code \citep[SME;][]{2017A&A...597A..16P}. For this study, the electron densities were derived with the assumption of non-LTE hydrogen ionisation by solving the statistical equilibrium equations imposing charge conservation \citep[][]{2007A&A...473..625L}.
\begin{figure}
   \centering

   \includegraphics[width=8cm]{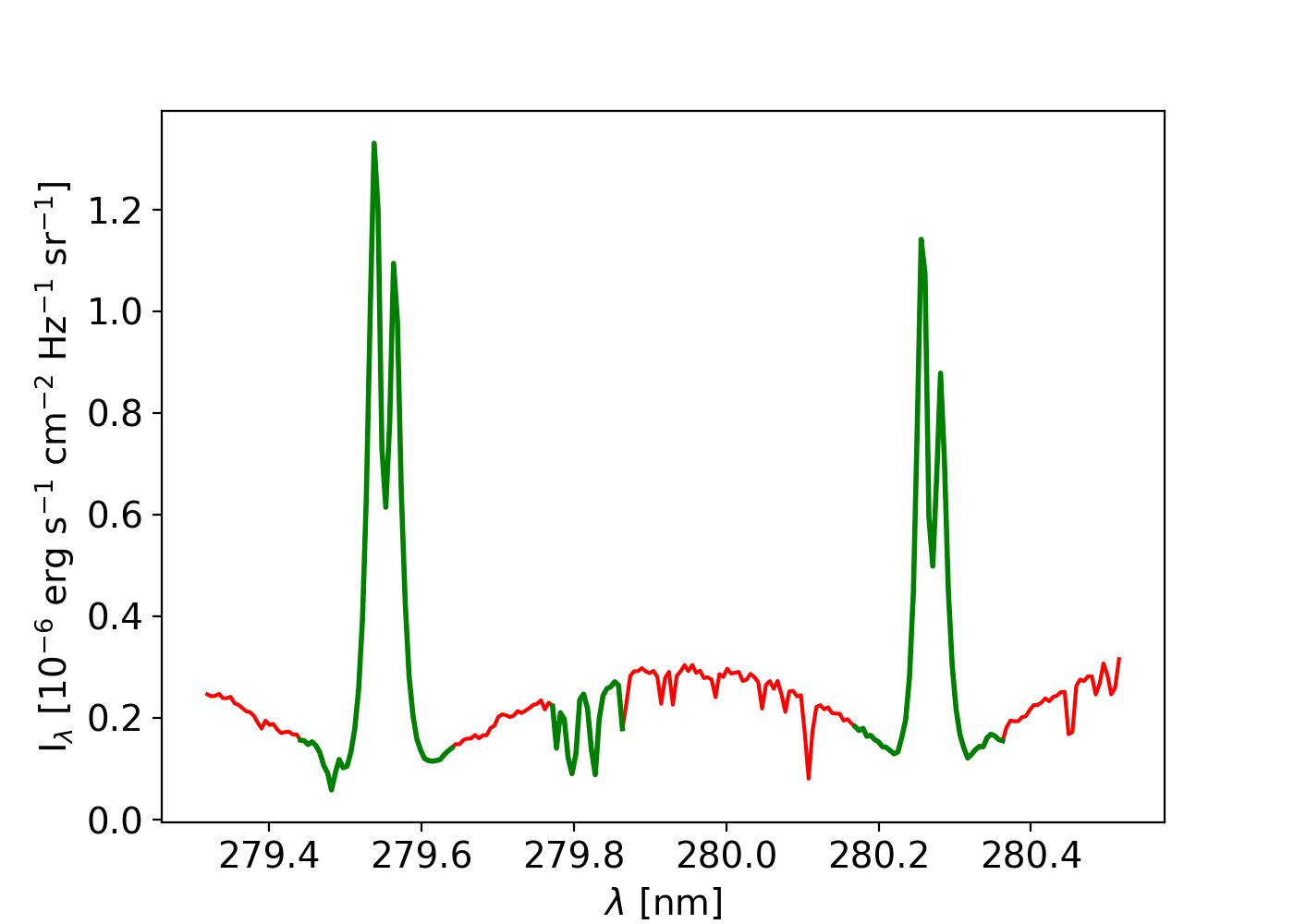}
   \caption{Typical \ion{Mg}{II} spectral line used for the inversions. The green line represents the spectral region that was used for the inversions. The spectral regions shown with red colour have been omitted to save computational time. This spectral profile corresponds to a dark fibril pixel (green star in Fig.~\ref{Figure:3}). }
              \label{Figure:4}%
\end{figure}  
    \begin{figure*}
   \centering

   \includegraphics[width=16.5cm]{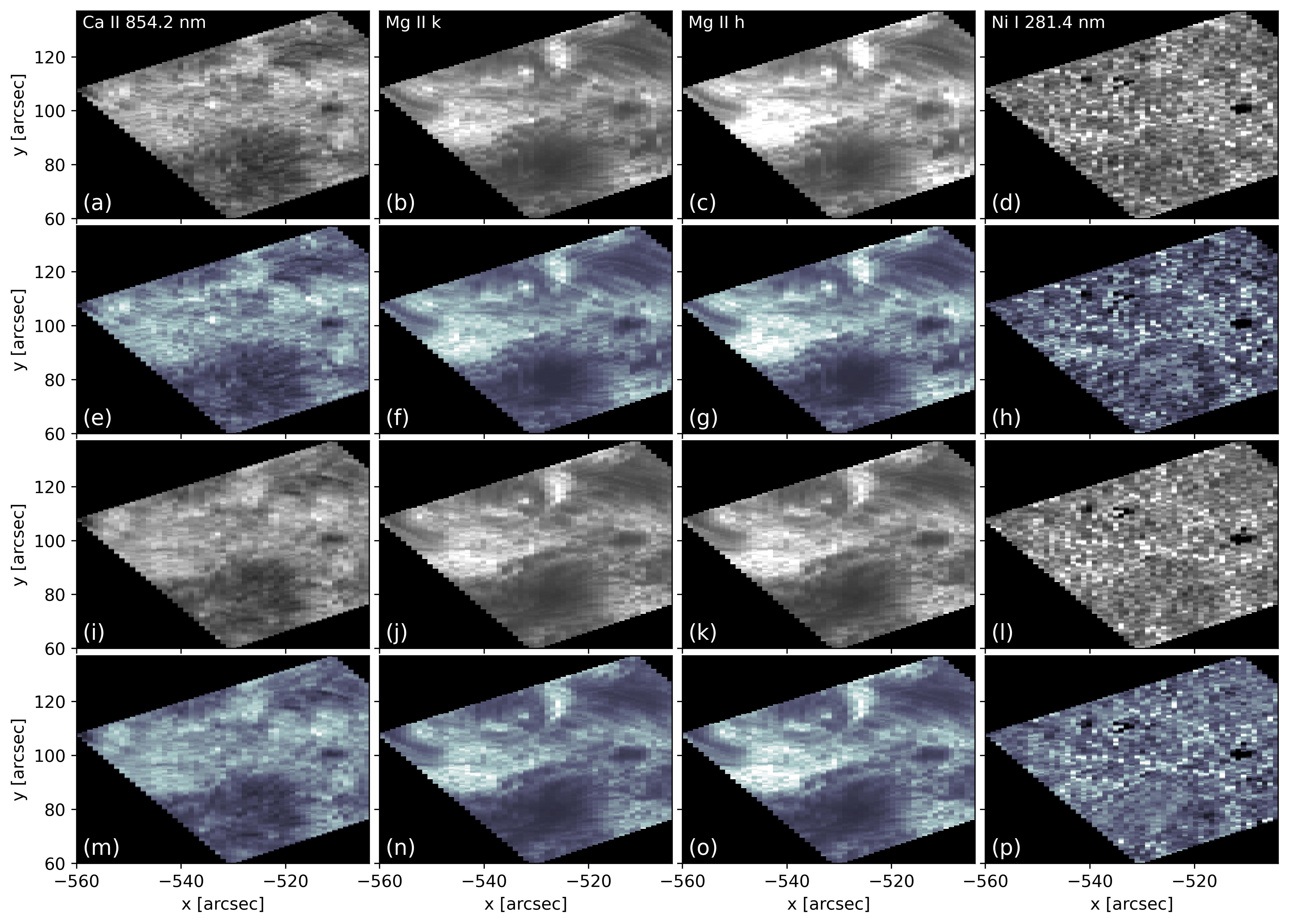}
   \caption{Comparison between the observed and inverted intensity maps. The grayscale maps represent the observations, while the blue intensity maps represent the inversion results. Panels a - d: Set A observations at different wavelength positions, the first at the core of the \ion{Ca}{II} line, the second and third columns at the cores of the \ion{Mg}{II} h \& k lines, and the fourth at the \ion{Ni}{I} 281.4~nm line core. Panels e - f: Intensity maps from the inversions of set A at the same wavelength positions. Panels i - l: Same as panels (a$)-$(d) for set B. Panels m - p: Same as panels (e$)-$(h) for set B. For comparison purposes, the intensity range for each pair of observed/inverted intensity maps is the same.}
              \label{Figure:5}%

\end{figure*}

\begin{figure*}
   \centering

   \includegraphics[width=17cm]{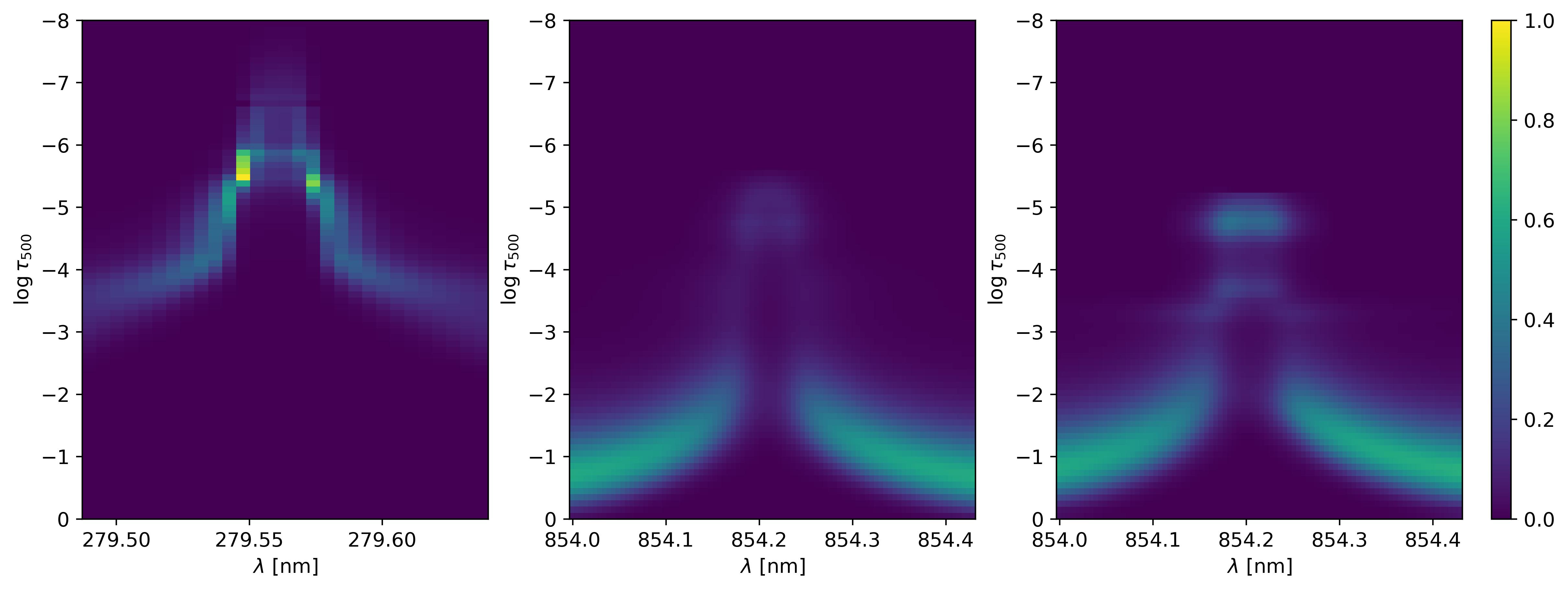}

   \caption{Response functions to variations in temperature. The first panel shows the average response function for the traced fibrils (see Sect.~\ref{sec:Fibril} and Fig.~\ref{Figure:8}) for the \ion{Mg}{II} k line, while the same is shown for the \ion{Ca}{II} 854.2~nm line in the middle panel. The panel on the right shows the response function to temperature for the \ion{Ca}{II} 854.2~nm for a representative pixel in a plage area, marked in panel a of Fig.~\ref{Figure:3}. All panels have a common normalisation factor, which is set to the peak value of the response function for the \ion{Mg}{II} k line in the first panel. }
              \label{Figure:6}%
\end{figure*}
It is important to note that the structures that were studied were mostly seen around the line cores of the \ion{Ca}{II} and \ion{Mg}{II} lines. The 1.5D plane parallel inversions performed with STiC cannot account for 3D radiative transfer effects that are important in these spectral regions \citep[][]{2009ApJ...694L.128L,2013ApJ...772...89L,2013ApJ...772...90L,2015ApJ...806...14P}.

  \subsection{Inversion set-up}

  Any given spectral line will be sensitive to different physical parameters across a limited range of heights in the atmosphere. Therefore, spectra coming from different pixels representative of different locations can look very similar if the properties of the atmosphere are not too dissimilar around the heights where the spectral line under consideration is formed. It is therefore logical to attempt a minimisation of the computation time spent in the inversions. With that in mind, a k-means algorithm was used to classify the nearly 17000 pixels into 500 groups, where the average spectral profile for each group was used. Therefore, 500 inversions (one for each group) were performed initially. This method made it possible to treat each of these 500 profiles individually, performing the inversion with modified initial parameters such as the number of nodes or the position of the transition region in the initial guess model for pixels where the initial inversion failed. Afterwards, for each individual pixel, the inverted atmosphere representative of its group is used to seed the inversion.

  A regular depth grid of 80 points spanning the range $\log \tau_{500} = [-8,0]$ (where $\tau_{500}$ is the optical thickness of the 500~nm radiation) was interpolated from a FAL-C \citep[][]{1985cdm..proc...67A,1993ApJ...406..319F} atmosphere and used as an initial guess for the inversions. In order to optimise the convergence of the inversions, the initial 500 profiles were inverted with a reduced number of nodes, and this number was increased for the subsequent inversion of individual pixels. Table~\ref{table:Table1} summarises the node distribution for the different inversion cycles. The number of nodes was slightly modified for failed pixels in order to properly invert them. Since no spectropolarimetric data were available, the only parameters inverted were the temperature, $T$, the line-of-sight component of the velocity, $v_{\mathrm{LOS}}$, and the microturbulent velocity, $v_{\mathrm{turb}}$.

  Atom models of \ion{Ca}{II} and \ion{Mg}{II} were used with 6 and 11 levels, respectively. The \ion{Ca}{II} 854.2~nm line was modelled with a Voigt profile, while the \ion{Mg}{II} h \& k lines were treated considering the effects of partial redistribution. Inside the IRIS near-ultraviolet window, the photospheric \ion{Ni}{I} line at 281.4~nm was also inverted, while considering local thermodynamic equilibrium in order to improve the response at photospheric heights \citep[][]{2013ApJ...778..143P}. The triplet lines of \ion{Mg}{II} that were also present in the spectral window were also inverted.
  
  In order to save on computation time, only the spectral region around the cores of the \ion{Mg}{II} h \& k lines along with the \ion{Mg}{II} triplet lines between them, were used for the inversions (see Fig.~\ref{Figure:4} for a typical inverted spectral profile). This method proved not only to be less time-consuming, but the cores of the \ion{Mg}{II} h \& k lines were also better fitted. As the \ion{Ni}{I} line is blended, only the spectral points around the line core were used.

\begin{figure*}
   \centering

   \includegraphics[width=16cm]{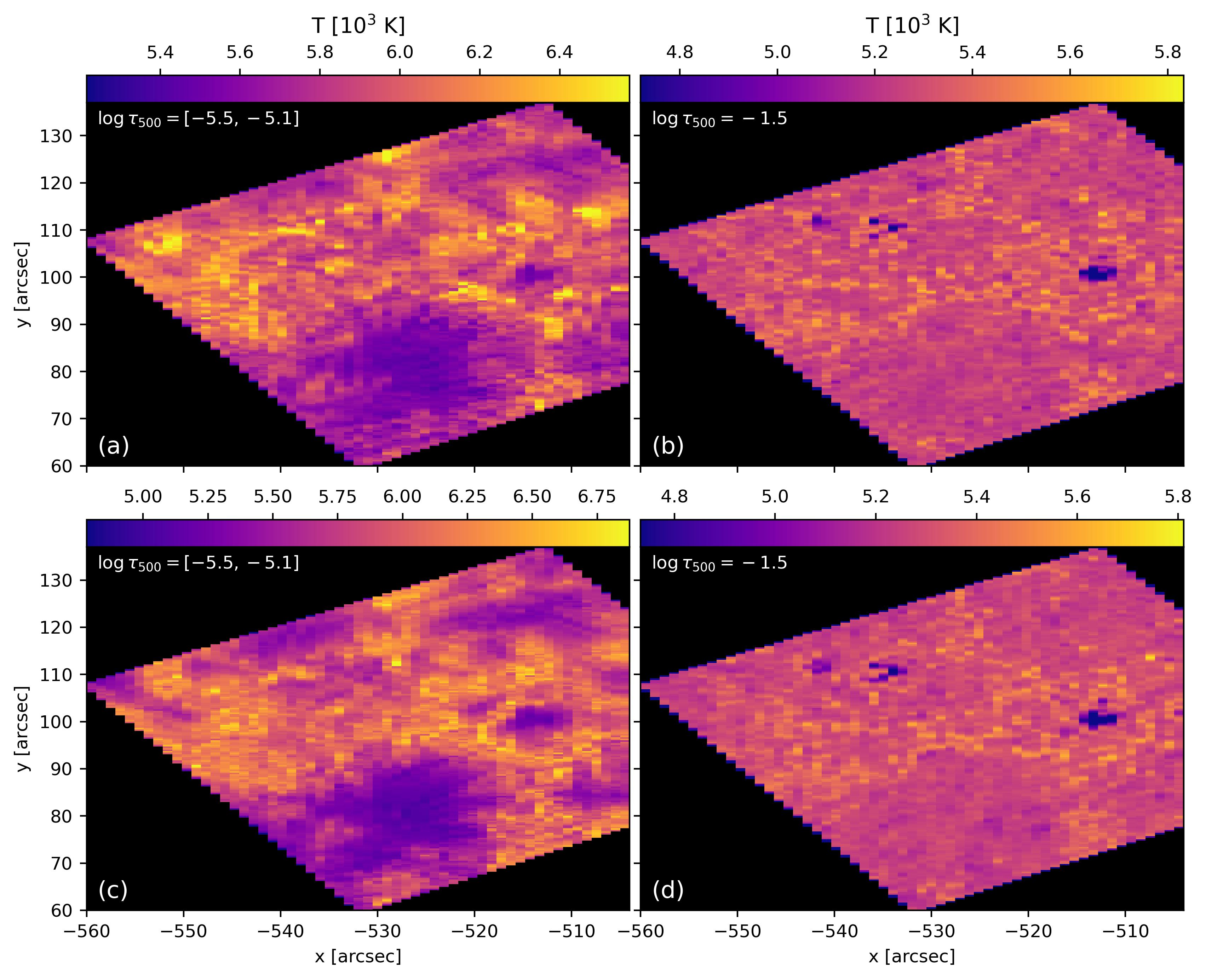}
   \caption{Results of the temperature inversions for (a) and (b) sets A and (c) and (d) set B.}
              \label{Figure:7}%
\end{figure*}

 \begin{figure*}
   \centering
   \includegraphics[width=19cm]{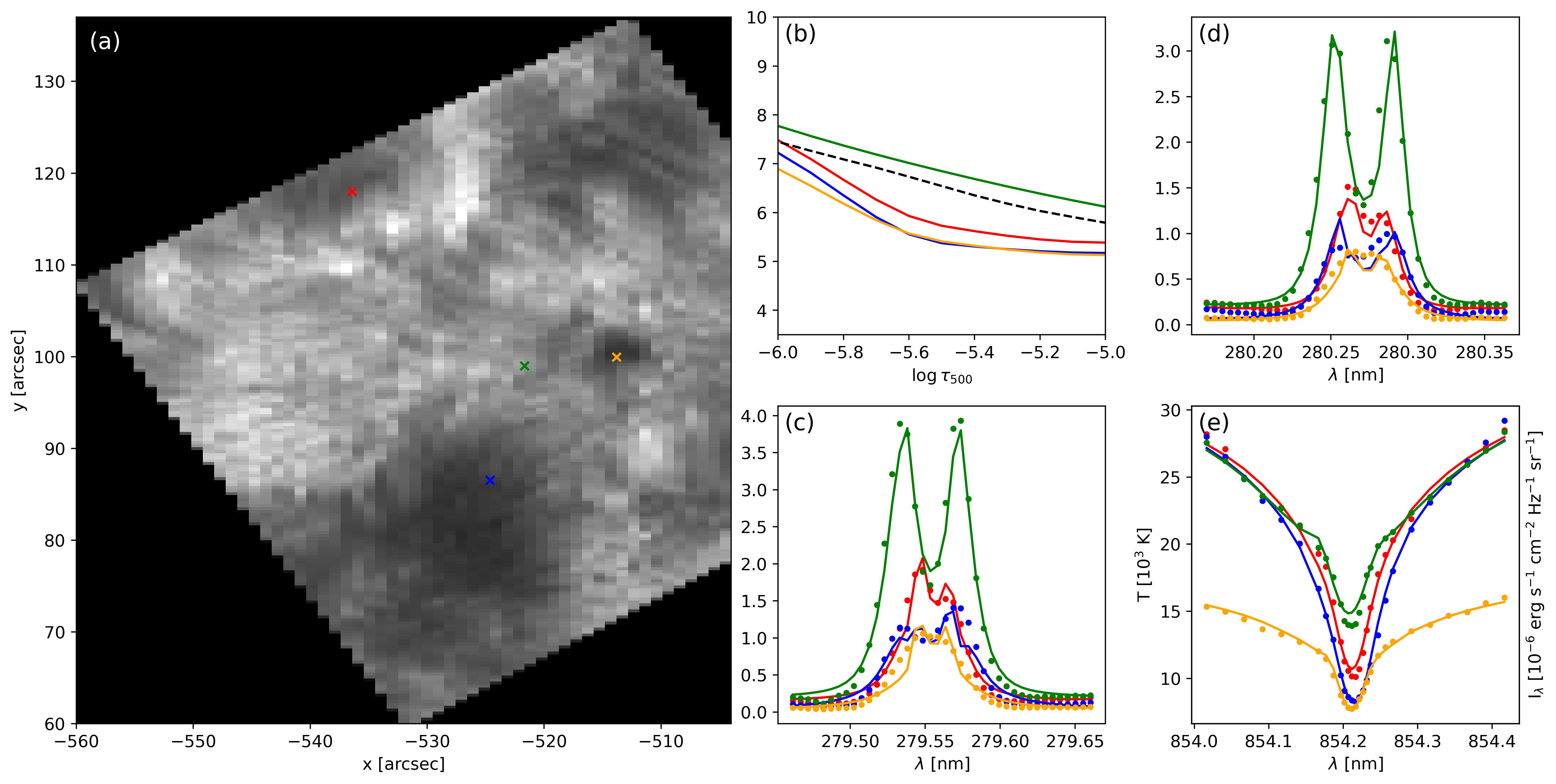}
   \caption{Examples of the inversion results for different pixels in the field of view of set B. (a) Intensity at the core of the \ion{Ca}{II} line. The colours of the marked pixels correspond to the examples displayed on the other panels. (b) Temperature stratification inversion results at the range $\log \tau_{500} = [-5, -6]$ for the different marked pixels on panel (a).  Black line corresponds to the temperature stratification of the original input model. (c) Synthesised profiles as a result of the inversions (continuous lines) compared to the observed spectral profiles (dots) for the \ion{Mg}{II} k line. (d) Same as panel (c) for the \ion{Mg}{II} h line. (e) Same as (c) and (d) for the \ion{Ca}{II} line. }
              \label{Figure:8}%
\end{figure*}
\begin{figure*}
   \centering

   \includegraphics[width=18cm]{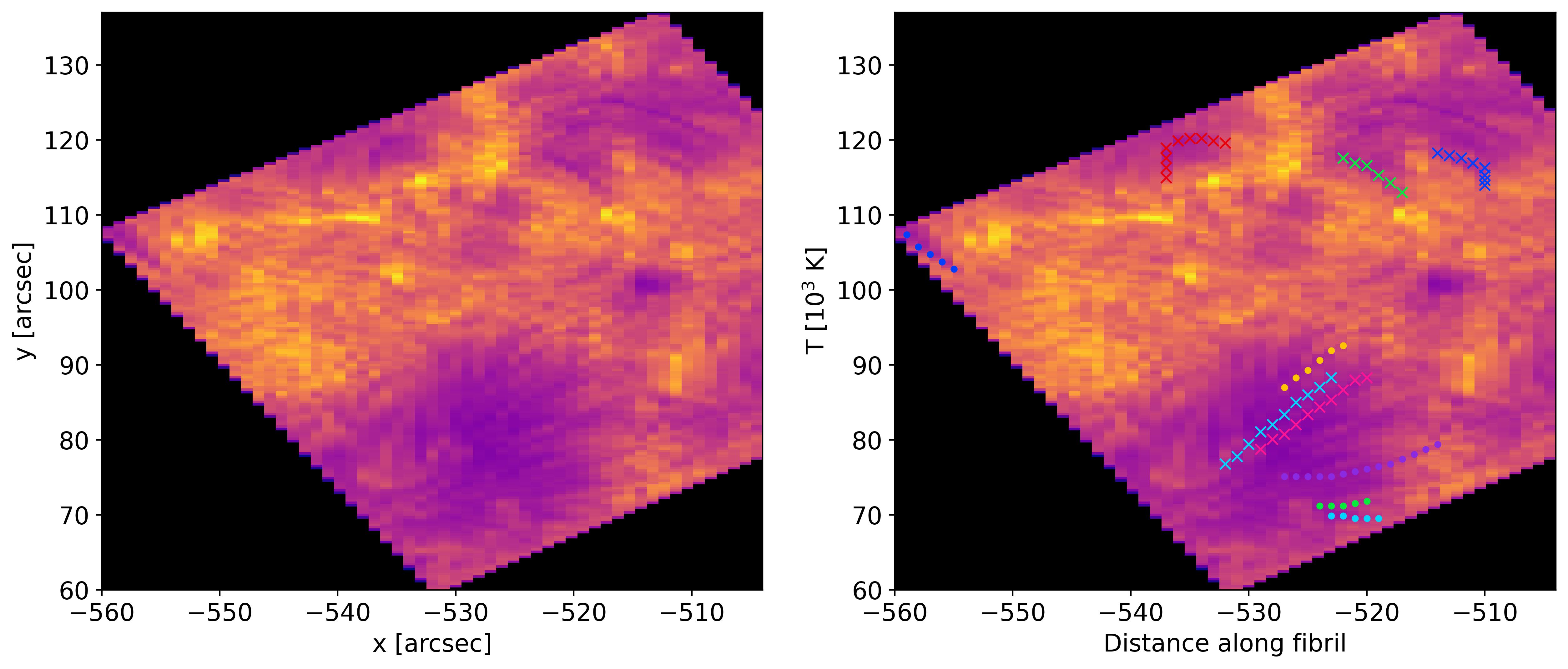}
   \caption{Example of the fibril tracing procedure. \textit{Left:} \ion{Ca}{II} line centre intensity image from set A. \textit{Right:}  Same image, with the traced fibrils overplotted. The style and colours of the plots match those described in Fig.~\ref{Figure:10}.}
              \label{Figure:9}%
\end{figure*}
 
\begin{figure*}
   \centering

   \includegraphics[width=18cm]{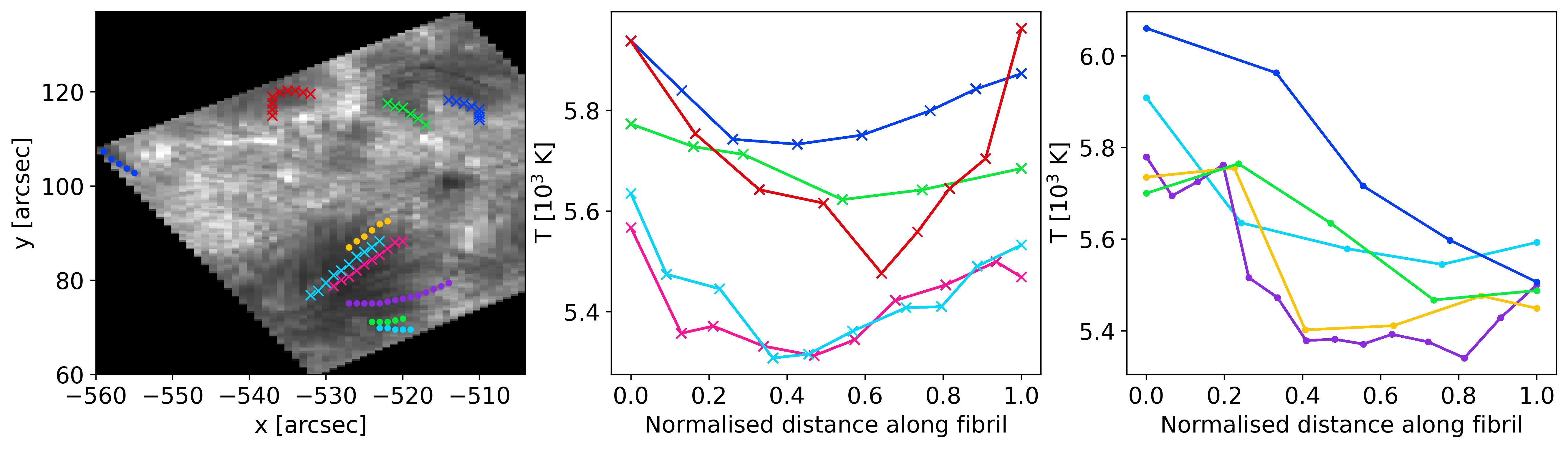}
   \includegraphics[width=18cm]{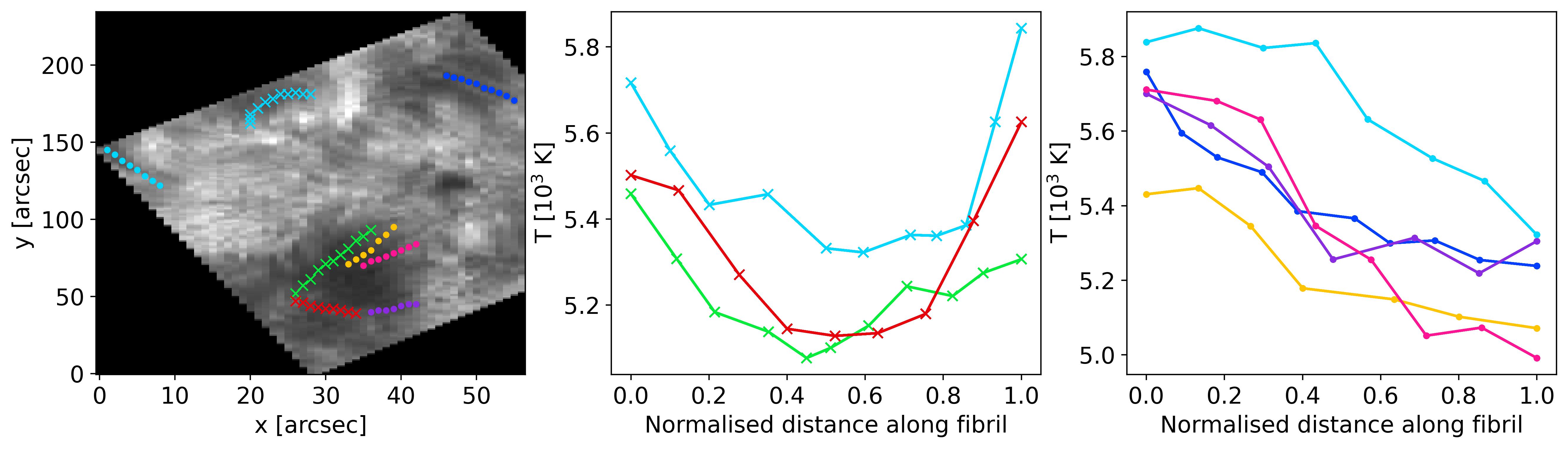}
   \caption{Results of the fibril tracing. The top and bottom panels display the results for sets A and B, respectively. The left panels display the outline of the traced fibrils over a \ion{Ca}{II} line centre image, with the colours matching that of the curves from the middle and right panels. Curves marked with an "x" correspond to group 1 fibrils, while dotted curves correspond to group 2 fibrils. The middle panels correspond to the average temperature at the range $\log \tau_{500} = [-5.1,-5.5]$ as a function of the position along the fibril for the group 1 fibrils. The plots on the right column show the same figures as the middle panels but for the fibrils of group 2. All the distances along the different fibrils have been normalised to the length of each fibril in order to simplify the plots.}
              \label{Figure:10}%
\end{figure*}
  \begin{table}
         \renewcommand*{\arraystretch}{1.5}
        \centering
        \caption{ Node distribution for each of the inversion cycles. Group and Individual refer to the number of nodes used for the inversion of the average profile of eack k-means group and for the inversion of each individual pixel, respectively.}
        \label{table:Table1}
        \begin{tabular}{ccccc}
        
            \hline \hline
            
            Parameter & \multicolumn{2}{c}{\ion{Ca}{II}} & \multicolumn{2}{c}{\ion{Ca}{II}, \ion{Mg}{II} and \ion{Ni}{I} }\\
            \cline{2-5}
            &Group&Individual&Group&Individual\\
            \hline
            $T$&7&9&9&11\\

            $v_{\mathrm{LOS}}$&2&4&3&4\\

            $v_{\mathrm{turb}}$&2&4&3&4\\\hline

        \end{tabular}
\end{table}

\section{Results and discussion} \label{results}

\subsection{Inversions}

The inversion results for sets A and B are displayed in Fig.~\ref{Figure:5}. The overall morphology of the inversions is in great agreement with the observations for all inverted lines. The intensity ranges were also nearly identical. The inversions were able to reproduce all the different structures present in the field of view, including the pores and granulation in the photosphere, the chromospheric fibrils, and the plage areas.

Spectral lines are not sensitive to the full range of optical depths within the solar atmosphere. Therefore, in order to determine where the inversion results actually come from within the atmosphere based on a given a model, response functions to the variation of physical parameters ought to be used. Response functions to changes in the model parameters can be performed numerically with STiC. The temperature response of the model showed that the  peak of the \ion{Ca}{II} line core response to temperature was, on average, found at $\log \tau_{500} = -5.2$ and at $\log \tau_{500} = -5.4$ for the emission peaks around the cores of the \ion{Mg}{II} lines (see Fig.~\ref{Figure:6}). Accordingly, the average temperature at the range $\log \tau_{500} = [-5.1,-5.5]$ is shown in Fig.~\ref{Figure:7} for both scans. This range of optical depths coincided with the locations where the temperature morphology resembled the observations at the line core of the lines the most. The temperature at $\log \tau_{500}=-1.5$ is also shown to represent the results obtained in the photosphere. Considering the limited spatial resolution, the results of the inversions were generally satisfactory. The photospheric pores and network were correctly resolved and the chromospheric slender dark structures were also recovered, albeit less sharply.

Some examples of the quality of the fits resulting from the inverted atmospheric model are displayed in Fig.~\ref{Figure:8}. When compared to the input model atmosphere, inverted temperatures at $\log \tau_{500} = [-5.1,-5.5]$ showed a slight temperature enhancement of about 300~K in the plage areas present in the data, while quiet Sun temperatures showed a diminution of up to $10^3$~K. This was an expected result, since the input model atmosphere is a result of the study of a larger area of the quiet solar atmosphere encompassing several different structures at the same time, giving a general view of how the solar atmosphere might be. Individual pixels, although of somewhat low spatial resolution, still represent much more specific areas of the atmosphere. The agreement between the observed and the inverted spectral profiles was better for the areas where the fibrils were located (blue and red crosses in Fig.~\ref{Figure:8}) than in the plage areas and the atmosphere above the photospheric pores (green and orange crosses, respectively). The inversion code in those areas especially struggled to fit the \ion{Ca}{II} line core data. However, since the focus of this study was to analyse fibrils, there was no further attempt to improve the fit results in plage areas. It is important to note that in some cases the central reversal region of the \ion{Mg}{II} lines were not reliably fitted. Radiative transfer calculations in three dimensions might be necessary in order to reproduce the central reversal region correctly \citep{2020ApJ...901...32J}. As the first panel of Fig.~\ref{Figure:6} suggests, the central reversal region of the \ion{Mg}{II} h \& k lines has a stronger response to temperature at higher atmospheric layers when compared to the emission peaks themselves, with a peak response near $\log \tau_{500} = -6$. This region is outside of the formation temperature of the fibrils analysed in this study.

\subsection{Fibril tracing} \label{sec:Fibril}

Despite the limiting nature of the data used in terms of spatial resolution and cadence, it was possible to manually trace the partial or complete outline of 18 different fibrils visible in the data (see an example for set A in Fig.~\ref{Figure:9}), divided into two groups: fibrils traced from footpoint to footpoint (group 1) and fibrils that were only partially traced from one footpoint until they were no longer detectable (group 2). The temperature variation in the range $\log \tau_{500} = [-5.1, -5.5]$ along the direction of the different fibrils for sets A and B is shown in Fig.~\ref{Figure:10}. The temperature along the different identified fibrils was not uniform; rather, it changed from a temperature maximum at the footpoints of the fibril, with a slight temperature decrease towards the midpoint  of the structure, while the temperature increased again nearly symmetrically towards the opposite footpoint. Group 2 fibrils showed at least part of this behaviour, with the temperature decreasing away from the footpoints. The inferred temperature at the footpoints of the structure was constrained between 5300~K and 6000~K, while the relative difference between the minimum temperature and the footpoint temperature was on the order of 10\%. One possible explanation for this very particular variation of the temperature along the chromospheric fibrils could be the signature of the heating mechanisms which they are subjected to, which perhaps is larger at the footpoints than at the apex of the fibrils. Recent studies have highlighted the increased heating present in the photospheric magnetic field concentrations, in particular in plage areas similar to the one studied in this work \citep[see][]{2021ApJ...921...39A,2022A&A...664A...8M}.

\begin{figure}
   \centering
   \includegraphics[width=8cm]{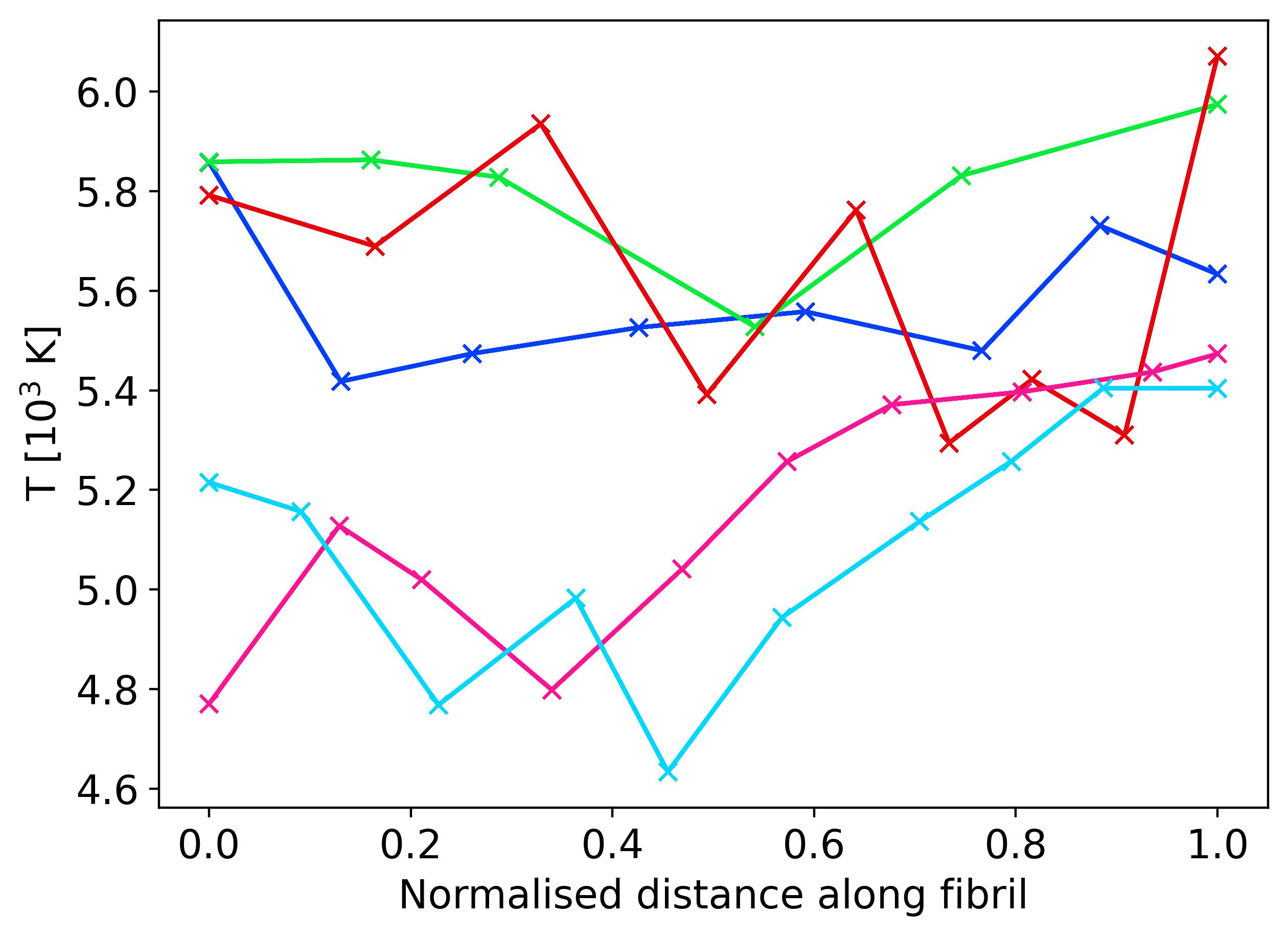}
   \includegraphics[width=8cm]{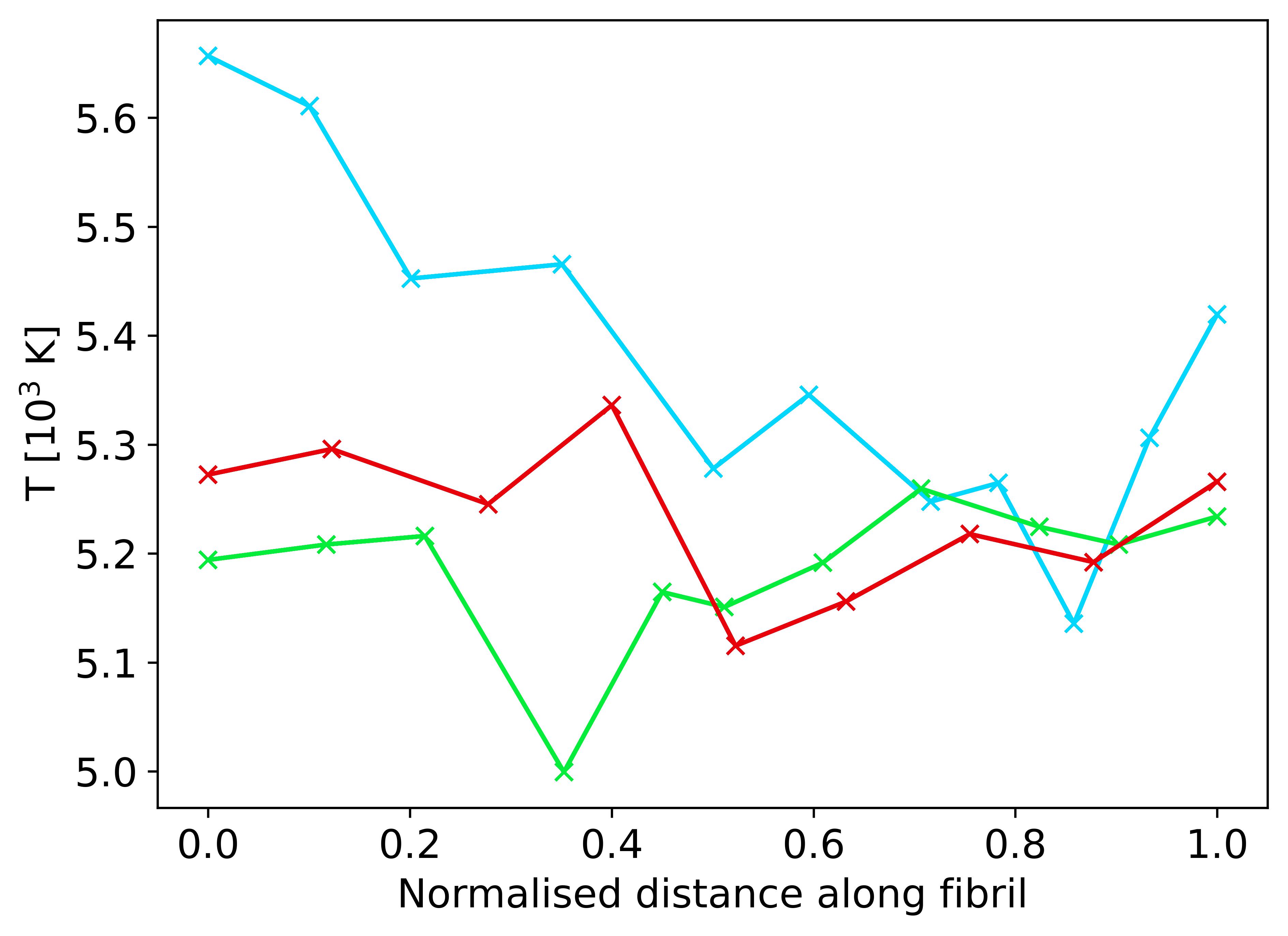}

   \caption{\ion{Ca}{II} inversions. \textit{Top}: Inferred temperature along the group 1 fibrils from set A. \textit{Bottom}: Same as the top panel, but for set B. Colours match those in the middle panels of Fig.~\ref{Figure:10}.  }
              \label{Figure:11}%
\end{figure}

\begin{figure}
   \centering
   \includegraphics[width=8cm]{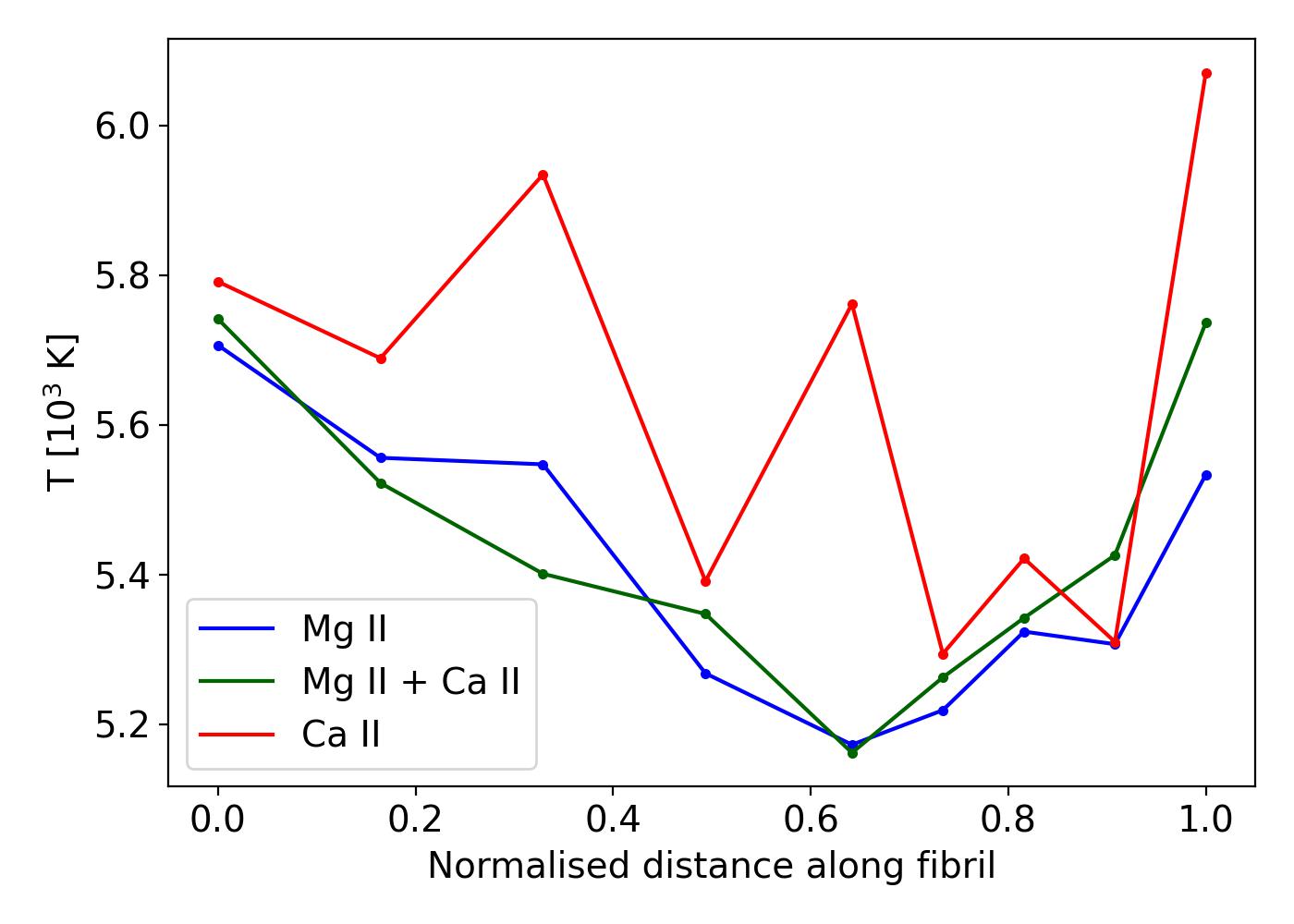}

   \caption{Multiline inversions. Comparison of the temperature variation along the traced fibril centred around ($X$, $Y$) = (-538\arcsec, 120\arcsec) from set A (see top-left panel of Fig.~\ref{Figure:10}) as inverted using the \ion{Ca}{II} data (red), the \ion{Mg}{II} data (blue), and the combined spectra from both elements (green).}
              \label{Figure:12}%
\end{figure}
The behaviour of the temperature along the different identified fibrils at the height where the response function to temperature attains its maximum value around the core of the \ion{Ca}{II} line was not seen when only the \ion{Ca}{II} line was inverted. Fig.~\ref{Figure:11} shows the temperature variation along the group 1 fibrils for both sets A and B from inversions considering only the \ion{Ca}{II} line. The inverted temperature along the fibrils remained rather uniform, albeit with some deviations for some of the fibrils. One reason for this discrepancy could be that the \ion{Ca}{II} line is not sufficiently sensitive to temperature to resolve the variation of temperature along fibrils inferred when information from the \ion{Mg}{II} lines is added. Additionally, the seeing conditions during the SST observations were not ideal, and the observations at the original resolution showed some atmosphere-induced effects that might have been partially mitigated when averaging the data spatially but nonetheless could still be affecting the quality of the spectral information\footnote{The full resolution SST image displayed in the left panel of Fig.~\ref{Figure:1} corresponds to the time of best seeing conditions at the SST. The conditions were quite variable during the rest of the observations}.

\subsection{Temperature sensitivity of the \ion{Mg}{II} h \& k and \ion{Ca}{II} 854.2 nm lines}
\begin{figure}
   \centering
    \includegraphics[width=8cm]{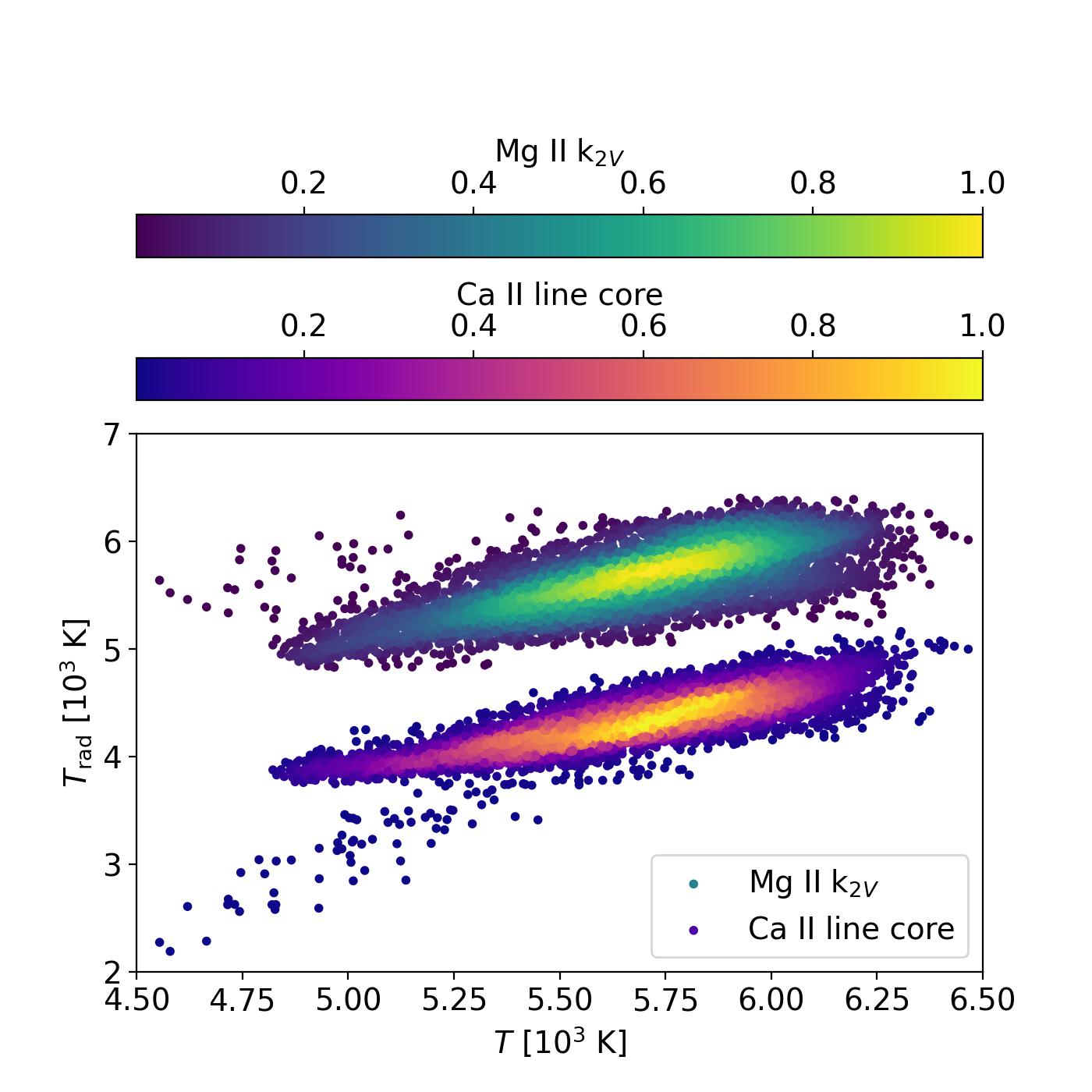}
    \includegraphics[width=8cm]{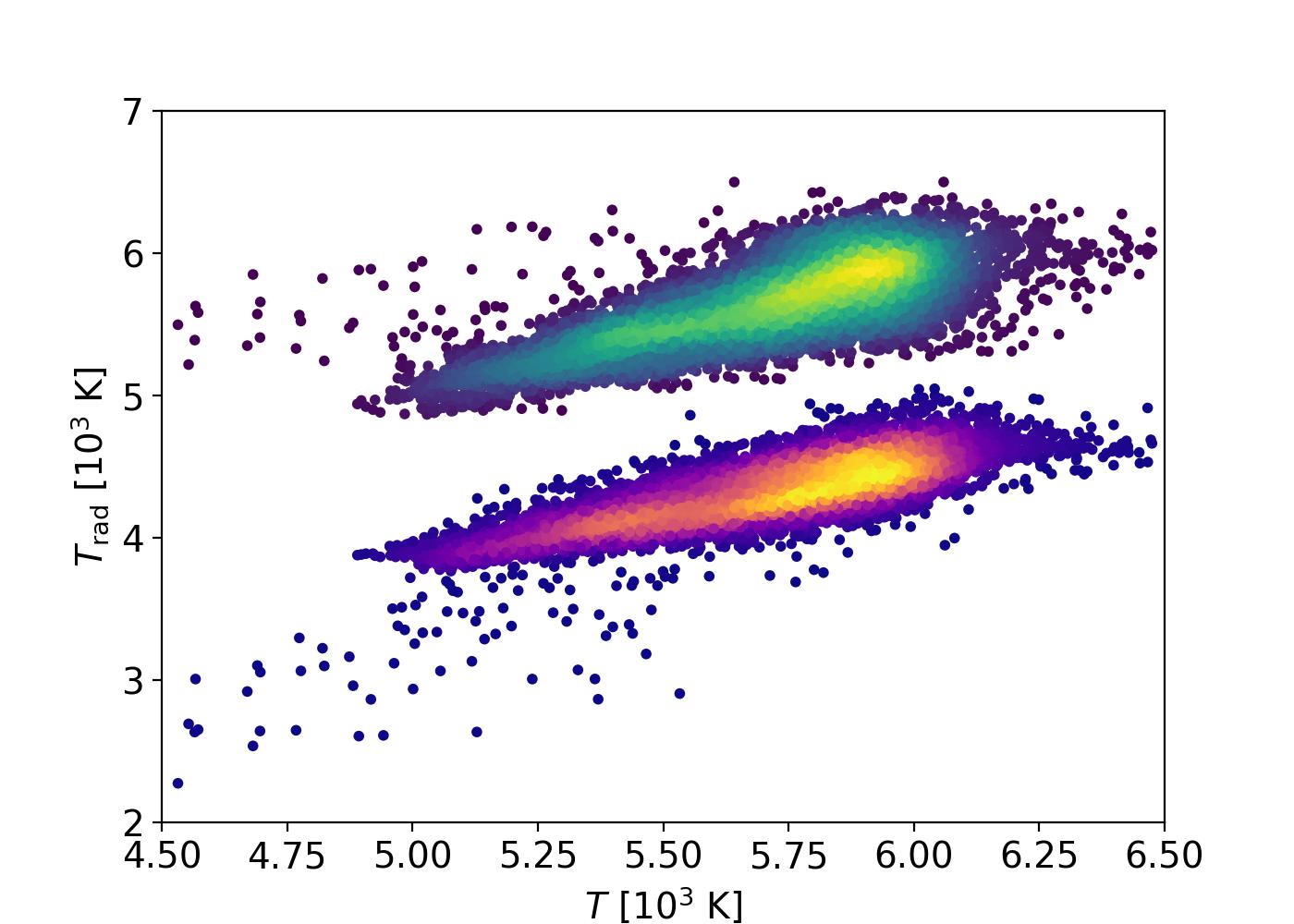}
   \caption{Radiation temperature vs inverted temperature. \textit{Top}: $T_\mathrm{rad}$ as a function of the temperature of the traced fibrils for set A, for the $\mathrm{k}_{2V}$ feature and the \ion{Ca}{II} line core. \textit{Bottom:}
    Same as top panel, for set B. The intensity of the dots plotted reflects the local density of points, normalised to unity.}
              \label{Figure:13}%
\end{figure}

As can be seen in Fig.~\ref{Figure:6}, the \ion{Ca}{II} line is sensitive to the temperature at the lower region of the range $\log \tau_{500} = [-5.1, -5.5]$, while the emission peaks and the line core of the \ion{Mg}{II} lines have their strongest response at the middle to upper part of this region. Figure~\ref{Figure:12} shows an example of the different temperature variation along a fibril as inverted using only the \ion{Ca}{II} line, only the \ion{Mg}{II} lines or the spectra from all lines combined. While the \ion{Mg}{II}-only inversions already have a similar shape to the inversions from the combined atomic species, the use of all lines seems to generate a more symmetric temperature variation, highlighting the contribution from all the lines to the final results. The fact that for some points along the fibril the inverted temperature using all atomic species differs by almost 600 K from that of the temperature inverted using only the \ion{Ca}{II} line could again be caused by the lower sensitivity of the intensity at the line core of the \ion{Ca}{II} line to temperature. There is a strong temperature sensitivity at the emission peaks of the \ion{Mg}{II} h \& k lines \citep[][]{2013ApJ...772...90L}. Defining $T_\mathrm{rad}$ as the radiation temperature obtained from inverting the Planck function for a given wavelength and observed intensity, Fig.~\ref{Figure:13} shows the variation of $T_\mathrm{rad}$ for the blue-shifted emission peak of the \ion{Mg}{II} k line \citep[$\mathrm{k}_{2V}$, see][]{2013ApJ...772...90L} and the line core of the \ion{Ca}{II} line as a function of the temperature of the traced fibrils from sets A and B. While there seems to be a near one-to-one (with noticeable spread) relation between the $T_\mathrm{rad}$ obtained from the $\mathrm{k}_{2V}$ feature and the inverted temperature, $T_\mathrm{rad}$ obtained from the \ion{Ca}{II} line core yields lower values, with an average temperature deficit of $1.3 \times 10^3$ K with respect to the inverted temperature. 

\subsection{Measuring $v_{\mathrm{LOS}}$ and $v_{\mathrm{turb}}$} 

Only four nodes were used in the final inversion cycle for the inference of $v_{\mathrm{LOS}}$ and $v_{\mathrm{turb}}$, as the main focus of this work was to study the temperature of chromospheric fibrils. Since the data analysed here is comprised mostly of low-lying, near-horizontal structures, the average inferred $v_{\mathrm{LOS}}$ for the identified fibrils was rather small, with an average upflow value of $-$1~km~$\mathrm{s}^{-1}$ for set A and a slight downflow of 0.3~km~$\mathrm{s}^{-1}$ for the set B. This could indicate that the motion of these structures was mostly constrained to the plane of the sky with respect to the observer.

The use of chromospheric lines from multiple atomic species in the inversions allowed us to ease the problem of the possible degeneracy that can appear when inferring temperatures and microturbulent velocities when only one species is used. The inverted microturbulence values along the different identified fibrils followed a similar behaviour to that of the temperature, with a maximum value at the footpoints, decreasing towards the center of the fibril. Maximum values ranged from 6 to 7~km~$\mathrm{s}^{-1}$, while at the center of the fibrils $v_{\mathrm{turb}}$ decreased to around 4~km~$\mathrm{s}^{-1}$.

\subsection{Comparison with previous works}
The temperatures inferred for both sets are in very good agreement with the non-LTE H$\alpha$ calculations of \citet[][]{1967AuJPh..20...81G,1967sp...conf..353G}, while being lower than the temperatures derived from cloud model calculations such as those performed by \citet[][]{1997A&A...324.1183T}. Cloud modelling has been widely used for the study of general physical properties in identifiable chromospheric structures, however, the non-LTE inversion scheme used here presents a significant step forward in terms of sophistication. Therefore, differences among the results are not unexpected. 

There are other types of fibrillar structures populating the chromosphere, such as the so-called \ion{Ca}{II} K bright fibrils \citep[][]{2009A&A...502..647P,2017ApJS..229...11J}. \citet[][]{2020A&A...637A...1K} used spectroscopic observations in the \ion{Ca}{II} K line, along with spectropolarimetric observations in the \ion{Ca}{II} 854.2~nm and \ion{Fe}{I} 630.2~nm lines obtained with the CRISP instrument to study the physical properties of these structures. They also performed inversions with STiC in order to study the thermodynamic and magnetic properties of bright \ion{Ca}{II} K fibrils. The results obtained in this work for dark chromospheric fibrils agree remarkably well with their temperature inversion results. These authors compared the appearance of the traced bright fibrils in different chromospheric lines, namely, the \ion{Ca}{II} K, the \ion{Ca}{II} 854.2~nm and the H$\alpha$ lines. They found that only in 14\% of the studied bright \ion{Ca}{II} K fibrils there is a correspondent co-temporal and co-aligned bright feature in the \ion{Ca}{II} 854.2~nm and the H$\alpha$ lines, with the vast majority of the cases corresponding to no bright fibrils in the other lines, bright fibrils in only one of the lines, or bright fibrils of a different appearance and orientation in the other lines. Here, it was found that the temperature near the photospheric magnetic field concentrations is larger than that of the midpoints of the fibrils, which is in agreement with the results from these authors. It has been suggested that the bright appearance of some chromospheric fibrils could be associated with temperature enhancements \citep[][]{2019ApJ...881...99M}. Indeed, \citet[][]{2020A&A...637A...1K} found that near large photospheric concentrations, the temperature of the head (here referred to as the 'footpoint') of the fibrils can be as high as 9000~K. However, these authors carried out a detailed study of a bright fibril that could be easily spotted in all three chromospheric lines, finding temperatures matching values reported here for dark fibrils. Therefore, while temperature enhancements might have an impact in the brightness of fibrils, there might also be additional physical mechanisms at play that would account for their bright or dark appearance.

\subsection{Spatial resolution effects}
A bright fibril is clearly seen in the left panel of Fig.~\ref{Figure:1} around ($X$, $Y$) = (-515\arcsec, 95\arcsec), however, it was not possible to detect it once the CRISP data had been spatially averaged and sampled at the positions of the IRIS slit. Although this inability to detect the structure was in part caused by the non-contiguous horizontal sampling of the raster observations, the spatial averaging could also blend the spectral signal from nearby structures into a single pixel. This is a problem that is especially significant at the boundaries of the fibrils.

The effect of spatial resolution can also cause changes to the inverted values of velocities. If the signal from structures with different $v_{\mathrm{LOS}}$ is mixed into a single pixel, the average spectral profile can become wider, leading the inversion code to interpret it as an effect of micro-turbulence.

\section{Conclusions} \label{conclusions}

In this paper, observations of a plage area obtained simultaneously with the CRISP instrument at the SST and the IRIS satellite were used to probe the thermodynamic properties of chromospheric fibrils. The CRISP data consisted of spectroscopic observations in the \ion{Ca}{II} 854.2~nm and in the H$\alpha$ lines. The observations obtained with IRIS consisted of raster scans in the \ion{Mg}{II} h \& k lines. The data from both observing stations were combined, aligned and calibrated in order to perform non-LTE inversions with the STiC inversion code. A k-means algorithm was employed in order to increase the speed of the inversions, based on the fact that spectral lines are sensitive to a limited range of heights in the solar atmosphere.

The manually detected fibrils present in the field of view were classified into two groups: a group of fibrils which were completely traced between both footpoints and a group of fibrils for which only one footpoint was unambiguously detected. The first group of fibrils showed a common variation of the average temperature at $\log \tau_{500} = [-5.1,-5.5],$ with the position along the fibril. They had a peak temperature at the footpoints which was nearly identical, while the temperature dropped nearly symmetrically towards the midpoint of the fibrils. The average difference in temperature between the maxima at the footpoints and the minimum at the midpoint was found to be (on average) 300~K. The partially resolved fibrils also showed a common behaviour, with a peak temperature at the footpoint that decreased towards the last detected point in the fibril.

While the variation of the temperature as a function of the position in the detected fibrils was determined from the inversions using all the different spectral lines available, when inverting only the \ion{Ca}{II} 854.2~nm line, this variation could not be reproduced. It could be a sign that the \ion{Ca}{II} 854.2~nm line alone does not have enough sensitivity to the temperature at chromospheric heights in order to completely resolve some of the structures present in this part of the solar atmosphere. The inclusion of the \ion{Mg}{II} h \& k lines, which have a very strong sensitivity to temperatures in the chromosphere, has proven to be essential in order to infer the temperature of fibrils. However, including IRIS data in the inversion meant that no time dependence of the temperature variation along the fibrils could be detected. Therefore, in order to study how the temperature structure might change if, for example, wave propagation takes place in the fibril, another observational set-up might be needed. A possibility that is left for future studies of the magnetism and thermodynamic structure of chromospheric fibrils is the combined use of spectropolarimetric observations of the \ion{Ca}{II} 854.2~nm line and the \ion{Ca}{II} H \& K lines, which can be simultaneously observed with filtergraphs from ground facilities such as the SST. However, this would require bearing in mind the possible degeneracy between the microturbulence and temperature that might remain unsolved since only one atomic species would be used for these inversions.

A future investigation of the differences between the dark and bright fibrils appears to be necessary in order to understand their properties and origin. The results obtained here for the temperature inversion of dark chromospheric fibrils agree very well with the inferred temperatures of bright \ion{Ca}{II} K fibrils of \citet[][]{2020A&A...637A...1K}. Adding the temperature sensitivity of the \ion{Mg}{II} h \& k lines could help to find a clear difference between these structures. It would also help to mitigate the degeneracy between microturbulence and temperature inherent of single-species studies. Giving that the spatial resolution of the data used here is significantly lower than the maximum attainable resolution with SST observations, a more thorough study of its effects and possible changes in the results obtained here with data of better seeing conditions quality is necessary.

\begin{acknowledgements}
This publication is part of the R+D+i project PID2020-112791GB-I00, financed by MCIN/AEI/10.13039/501100011033. MK acknowledges the support from the Vicepresidència i Conselleria d’Innovació, Recerca i Turisme del Govern de les Illes Balears and the Fons Social Europeu 2014-2020 de les Illes Balears. This research has made use of SunPy v1.1, an open-source and free community-developed solar data analysis Python package \citep{SunPy2020}. The Swedish 1-m Solar Telescope is operated on the island of La Palma by the Institute for Solar Physics of Stockholm University in the Spanish Observatorio del Roque de los Muchachos of the Instituto de Astrof\'\i sica de Canarias. The Institute for Solar Physics is supported by a grant for research infrastructures of national importance from the Swedish Research Council (registration number 2017-00625). We also thank P. Antolin and N. Freij for the data acquisition. We thank the anonymous referee for her/his comments, that helped improve the manuscript. 
\end{acknowledgements}

\bibliographystyle{aa}
\bibliography{citations}

\end{document}